\documentclass[apj]{emulateapj}
\usepackage{apjfonts}
\usepackage{amsbsy}

\usepackage{color}

\def\hmpc{$h^{-1}$Mpc}
\def\hkpc{$h^{-1}$kpc}

\def\mstar{$M_\ast$}

\def\hmsol{$h^{-1}$M$_\odot$}

\def\navg{\langle N\rangle_M}

\def\nsat{\langle N_{\rm sat}\rangle_M}
\def\ncen{\langle N_{\rm cen}\rangle_M}
\def\mmin{M_{\rm min}}
\def\mcut{M_{\rm cut}}

\def\asat{\alpha_{\rm sat}}

\def\om{\Omega_m}
\def\oml{\Omega_\Lambda}
\def\omb{\Omega_b}
\def\s8{\sigma_8}

\def\lcdm{$\Lambda$CDM}

\def\x2{$\chi^2$}
\def\hmsol{$h^{-1}\,$M$_\odot$}

\def\NNm1{\langle N(N-1) \rangle}

\def\dc{\delta_c}

\def\fsat{f_{\rm sat}}

\def\m_star{M_\ast}

\def\lcdm{$\Lambda$CDM}

\def\om{\Omega_m}

\def\omb{\Omega_b}
\def\s8{\sigma_8}

\def\hmpc{$h^{-1}\,$Mpc}
\def\hkpc{$h^{-1}\,$kpc}

\def\x2{$\chi^2$}
\def\hmsol{$h^{-1}\,$M$_\odot$}

\def\mstar{M_\ast}
\def\mmin{M_{\rm min}}

\def\mcut{M_{\rm cut}}

\def\navg{\langle N\rangle_M}

\def\nsat{\langle N_{\mbox{\scriptsize sat}}\rangle_M}

\def\ncen{\langle N_{\mbox{\scriptsize cen}}\rangle_M}

\def\NNm1{\langle N(N-1) \rangle}

\def\dc{\delta_c}

\def\fsat{f_{\rm sat}}

\def\dc{\delta_c}

\def\fsat{f_{\rm sat}}

\def\mstar{M_\ast}

\def\p0{P_0(r)}

\def\m12{M_{12}}

\def\m12{M_{12}}

\def\rs0{\hat{R}_{\rm sh}^0}
\def\mg2{Mg\,II}

\def\lstar{L_\ast}

\def\lfdrg{\Phi_{\rm DRG}}
\def\mp{\mu_{\rm cen}}
\def\fcen{f_{R\rm cen}}
\def\fmax{f_{R\rm max}}
\def\fsat{f_{R\rm sat}}
\def\hcubed{(h^{-1}\,{\rm Mpc})^{-3}}

\def\wth{w(\theta)}
\def\hgpc{h^{-1}\,{\rm Gpc}}
\def\ndrg{\bar{n}_{\rm DRG}}
\def\xics{\xi_{\rm 1h}^{cs}(r)}
\def\xiss{\xi_{\rm 1h}^{ss}(r)}

\def\ndrg{\bar{n}_{\rm DRG}}

\begin{document}

\title{Interpreting the Clustering of Distant Red Galaxies}

\author{ Jeremy L. Tinker$^1$, Risa H. Wechsler$^2$, \& Zheng
  Zheng$^3$ } \affil{$^1$ Berkeley Center for Cosmological Physics,
  University of
  California-Berkeley; {\tt tinker@berkeley.edu}\\
  $^2$ Kavli Institute for Particle Astrophysics and Cosmology, Dept. of Physics, and SLAC National Laboratory, Stanford University\\
  $^3$John Bahcall Fellow, School of Natural Sciences, Institute for
  Advanced Study}

\begin{abstract}

  We analyze the angular clustering of $z\sim 2.3$ distant red
  galaxies (DRGs) measured by \cite{quadri_etal:08}. We find that,
  with robust estimates of the measurement errors and realistic halo
  occupation distribution modeling, the measured clustering can be
  well fit within standard halo occupation models, in contrast to
  previous results. However, in order to fit the strong break in
  $\wth$ at $\theta=10^{\prime\prime}$, nearly all satellite galaxies
  in the DRG luminosity range are required to be DRGs. Within this
  luminosity-threshold sample, the fraction of galaxies that are DRGs
  is $\sim 44\%$, implying that the formation of DRGs is more
  efficient for satellite galaxies than for central galaxies. Despite
  the evolved stellar populations contained within DRGs at $z=2.3$,
  90\% of satellite galaxies in the DRG luminosity range have been
  accreted within 500 Myr. Thus, satellite DRGs must have known they
  would become satellites well before the time of their
  accretion. This implies that the formation of DRGs correlates with
  large-scale environment at fixed halo mass, although the large-scale
  bias of DRGs can be well fit without such assumptions. Further data
  are required to resolve this issue. Using the observational estimate
  that $\sim 30\%$ of DRGs have no ongoing star formation, we infer a
  timescale for star formation quenching for satellite galaxies of
  $450$ Myr, although the uncertainty on this number is large.
  However, unless all non-star forming satellite DRGs were quenched
  before accretion, the quenching timescale is significantly shorter
  than $z\sim 0$ estimates. Down to the completeness limit of the
  Quadri et al sample, we find that the halo masses of central DRGs
  are $\sim 50\%$ higher than non-DRGs in the same luminosity range,
  but at the highest halo masses the central galaxies are DRGs only
  $\sim 2/3$ of the time.
\end{abstract}


\section{Introduction}

Distant red galaxies (hereafter DRGs) are bright
near-infrared-selected galaxies that have been detected out to $z\sim
3$ (\citealt{franx_etal:03, van_dokkum_etal:03, daddi_etal:03}). These
galaxies are massive objects containing a significant population of
evolved stars (\citealt{forster_etal:04}), with many galaxies having
highly suppressed star formation or possibly none at all (\citealt{labbe_etal:05,
  kriek_etal:06}). Given the remarkable fact that these evolved (and
possibly dead) galaxies existed a mere 2 Gyr after the big bang, many
studies have further sought to measure their abundance and clustering
to determine their connection to dark matter structure
(\citealt{daddi_etal:03, grazian_etal:06, quadri_etal:07,
  quadri_etal:08, marchesini_etal:07, ichikawa_etal:07}).

\cite{daddi_etal:03} measured a very strong clustering strength for
DRGs, implying that these galaxies live in only the most massive dark
matter halos that exist at such high redshift. In fact, the number
density of such halos was significantly smaller than the observed
space density of the DRGs themselves. \cite{daddi_etal:03} used
measurements of small-scale clustering to find the correlation length
of their sample through extrapolation to larger scales. Using the halo
occupation distribution to model the clustering (HOD; see
\citealt{cooray_sheth:02} for a review), \cite{zheng:04} demonstrated
that the high clustering strength and observed number density were
compatible by taking into account a small fraction of DRGs that are
satellite galaxies orbiting around other DRGs that sit at the center
of the dark matter halo. This small satellite fraction produces a
significant difference in the slope of the correlation function at
large and small scales.

\cite{quadri_etal:08} (hereafter Q08) used a much larger sample of
DRGs to confirm the strong clustering of \cite{daddi_etal:03} and
subsequent studies, both at large ($\sim 20$ \hmpc) and small ($\sim
200$ \hkpc) scales. They attempted to model the data with a somewhat
different HOD approach from \cite{zheng:04}, but they could not find a
satisfactory fit: models that fit at small scales underpredicted the
large-scale bias, while models that reproduced the large-scale bias
overpredicted the small-scale clustering. They proposed that standard
assumptions of HOD models do not describe these galaxies; their halo
occupation correlates with large-scale environment as well as halo
mass, i.e., the so-called ``assembly bias'' \citep{croton_etal:07,
wechsler_etal:06, gao_white:07}. This correlation would imply that
galaxy formation depends on large-scale environment at fixed halo
mass, a correlation not seen in low-redshift data
(\citealt{tinker_etal:08_voids}).

In this paper we demonstrate that the observations of Q08 can be
well-fit by standard HOD models. The inclusion of cosmic variance in
the large-scale clustering errors ameliorates much of the discrepancy
with their model fits. We also show that a more physically motivated
halo occupation model yields significantly better results in fitting
the data as well. Further, we will use our HOD modeling to test the
mechanisms for the formation of red galaxies at $z>2$.

Unless otherwise stated, all calculations adopt a flat \lcdm\
cosmology consistent with the latest constraints from CMB anisotropies
(\citealt{dunkley_etal:08}). Our cosmological parameter set is
$(\om,\s8,h,n_s,\omb) = (0.25, 0.8, 0.7, 0.95, 0.045)$. All distances
are comoving. Due to their color selection, throughout this paper we
will use the terms DRGs and ``red galaxies'' interchangeably.

\begin{figure*}
\epsscale{1.0} 
\plotone{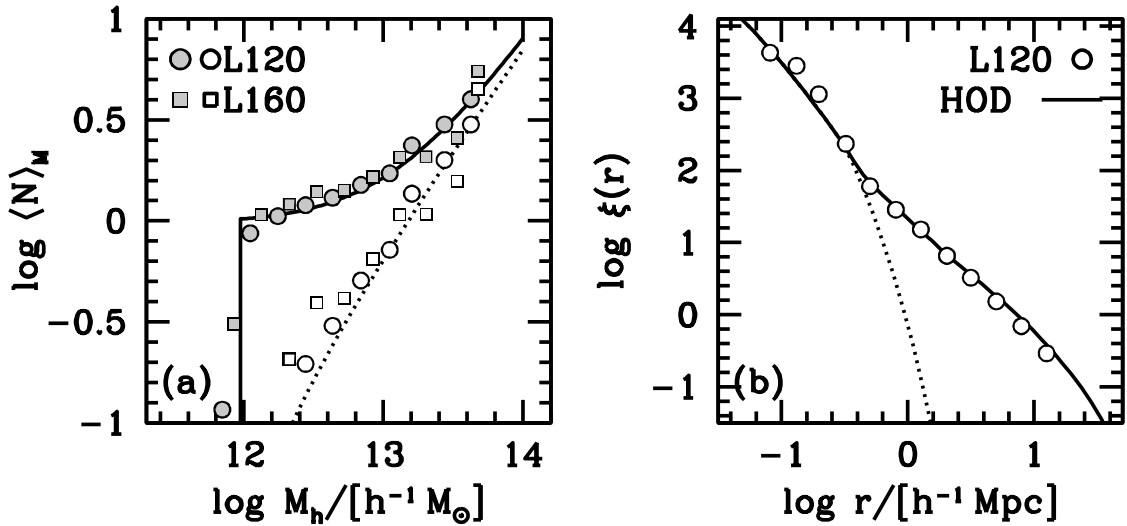}
\vspace{-7.5cm}
\caption{ \label{hod} Panel (a): Halo occupation functions produced by
  the abundance matching method described in the text. Gray symbols
  represent all halos (parent+sub) while white symbols represent only
  the subhalos. The two simulations have two different cosmologies, so
  they have been shifted to a common mass scale to demonstrate
  self-similarity. The solid and dotted curves represent the HOD used
  in all analytic calculations henceforth. Panel (b): Comparison
  between the clustering in the L120 box and the analytic model. The
  HOD for the analytic model is obtained from fitting the simulation
  results (e.g., panel a, but shifted to the proper mass scale). The
  solid line is the full correlation function, and the dotted line the
  one-halo term.}
\end{figure*}

\section{Methods}

\subsection{Definition of the Term ``Halo''}

Because we will be dealing with halos that exist in various
environments, it is important to have a clear definition of a halo and
discuss what it implies. We assume that all galaxies live at the
center of a virialized clump of dark matter. That dark matter clump
may be isolated or it may exist within the virial radius of a larger
structure. Therefore we will use the term {\it halo} to refer to an
object that is distinct; i.e., it does not exist within the virial
radius of another object. These objects typically have a mean
overdensity of $\sim$200 times that of the background universe. We
refer to objects inside the virial radius of halos as {\it
  subhalos}. We use the term {\it galactic halos} to refer to all
halos, both halos and subhalos, that contain galaxies at their center.

\subsection{Halo Occupation from Simulations}

Although collisionless $N$-body simulations do not include any baryon
physics, one can associate the likely sites of galaxy formation with
the dark matter halos and subhalos within a simulation.  Several
recent studies have demonstrated the robustness of this assumption by
comparing the clustering of galaxies to that of a sample of galactic
halos with the same space density; i.e., galaxies brighter than a given
luminosity threshold compared with galactic halos more massive than a
threshold that yields the same abundance.  \cite{conroy_etal:06} found
that the predicted galactic halo clustering was consistent with galaxy
two-point clustering measurements from $z=0$ to $z=5$ (see also
\citealt{kravtsov_etal:04, wang_etal:06}). \cite{marin_etal:08} extended
this to measurements of the galaxy three-point correlation function as
well.

For the purpose of this paper, we use high-resolution cosmological
$N$-body simulations to guide our choice of the halo occupation of all
galaxies (DRGs and non-DRGs). The HOD for all galaxies is then fixed
by making use of the luminosity function of all galaxies (see below),
and we focus our effort on constraining the DRG HOD from the
clustering data.

Just knowing where the galaxies are, however, doesn't identify which
ones are red. Before creating a model for the halo occupation of DRGs,
we first use the subhalo abundance matching technique (SHAM) to set
the occupation of {\it all} galaxies down to the completeness limit of
Q08. The space density of DRGs in the Q08 sample is $\ndrg=6.5\times
10^{-4}$ $\hcubed$ down to their completeness limit of $K<21$. Using
the $z\sim 2.3$ luminosity function of \cite{marchesini_etal:07}, the
space density of all galaxies is $1.5\times 10^{-3}$ $\hcubed$ at the
same magnitude threshold, $M_R=-22.3$, yielding a DRG fraction of
$44\%$.\footnote{As Q08 point out, the space density of DRGs from
  \cite{marchesini_etal:07} is slightly lower than that of the larger
  Q08 sample. Thus, whenever using the \cite{marchesini_etal:07}
  luminosity functions and data, a correction factor of $6.5/5\approx
  1.3$ is applied. This increases the published number density of
  galaxies brighter than $M_R=-22.3$ in \cite{marchesini_etal:07} from
  $1.2\times 10^{-3}$ to $1.5\times 10^{-3}$ $\hcubed$.} Whenever
referring to the sample of {\it all galaxies}, we mean all galaxies
(DRGs and non-DRGs) down to the completeness limit of the Q08 sample.


\begin{deluxetable}{cccc}
\tablecolumns{4} 
\tablewidth{20pc} 
\tablecaption{List of Simulations} 
\tablehead{ \colhead{$L_{\rm box}$ (\hmpc)} & \colhead{$(\om, \s8, n_s)$} & \colhead{ $m_p$ [\hmsol]} & \colhead{$z_{\rm out}$} }

\startdata

120 & $(0.3, 0.9, 1.0)$ & $1.07\times 10^9$ & 2.0 \\
160 & $(0.24, 0.75, 0.95)$ & $2.54\times 10^8$ & 2.5 \\
1000 & $(0.27, 0.8, 0.95)$ & $6.98\times 10^{10}$ & 2.5 \\

\enddata
\tablecomments{Each simulation will be referred to in the text by its
  box size. All simulations were performed with the ART code of
  \cite{kravtsov_etal:97}. The L120 and L1000 simulations have been
  described in \cite{tinker_etal:08_mf}.}
\end{deluxetable} 

Figure \ref{hod}a shows the halo occupation functions, $\navg$, of
halos and subhalos above a given maximum circular velocity such that
the space density of all galaxies down to the Q08 limit is
obtained. We use results from the first two simulations listed in
Table 1. Halos and subhalos are identified in the simulation by the
algorithm described in \cite{kravtsov_etal:04}, a variant of the
spherical overdensity halo finder. Because the simulations have
different cosmologies and redshifts, $\navg$ will differ between the
two. However, the {\it shape} of $\navg$ is self-similar. The HOD from
both simulations is well approximated by a satellite occupation
function of the form

\begin{equation}
\label{e.nsat}
\nsat = \left(\frac{M}{M_1}\right)^{\asat}\exp\left(-\frac{\mcut}{M}\right),
\end{equation}

\noindent where $\asat=1$, and a central occupation function of the form 

\begin{equation}
\label{e.ncen}
\ncen = \left\{ \begin{array}{ll}
  1 & {\rm if\ \ } M\ge \mmin \\
  0 & {\rm if\ \ } M< \mmin.\\
\end{array}
\right. 
\end{equation}

\noindent Both simulations exhibit the mass ratios $M_1/\mmin=15.7$
and $\mcut/\mmin=1.14$. To match the number density of all galaxies in
our fiducial cosmology, a value of $\mmin=9\times 10^{11}$ \hmsol\ is
required. This yields values of $M_1=1.4\times 10^{13}$ \hmsol, and
$\mcut=1.0\times 10^{12}$ \hmsol. In Figure \ref{hod}a, the HOD from
L120, a simulation with WMAP1 cosmological parameters
(\citealt{spergel_etal:03}), has been shifted by $-0.25$ dex in halo
mass. The results from L160, a simulation closer to the WMAP3
cosmological parameter set (\citealt{spergel_etal:07}), have been
shifted by $+0.23$ dex. These shifts bring the HODs into alignment
with the HOD for our fiducial cosmology. In the L120 box at $z=2$, the
total fraction of galactic halos that are subhalos is 13\%.  This is
in good agreement with \cite{conroy_etal:06} but somewhat smaller than
the subhalo fraction of \cite{wetzel_etal:08}, who find 18\%. This is
most likely attributable to differing classifications of a subhalo. In
our definition, a halo becomes a subhalo when it passes the spherical
virial radius of a larger halo. In \cite{wetzel_etal:08}, halos become
subhalos when they are linked by the friends-of-friends algorithm,
which can link objects outside of what we have defined as the virial
radius.

With $\navg$ specified, we can analytically calculate the galaxy
two-point autocorrelation $\xi(r)$. Our model for $\xi(r)$ is based on
the analytic model detailed in \cite{zheng:04} and
\cite{tinker_etal:05}. We use a Poisson distribution for the scatter
about $\nsat$, in agreement with simulations and semi-analytic models
(\citealt{kravtsov_etal:04, zheng_etal:05}) and with observations of
galaxy clusters (\citealt{kochanek_etal:03, lin_etal:04}). We also
assume that the satellite galaxies follow the radial distribution of
dark matter within the halos (e.g., the density profile of
\citealt{nfw:96}), using the halo concentrations of
\citealt{zhao_etal:08}. We use the \cite{tinker_etal:08_mf} mass
function for $\Delta=200$ and a new halo bias function based on those
simulations (Tinker et.~al., in prep; see Appendix A). Our analytic
model incorporates scale-dependent halo bias and halo exclusion for
proper modeling of the transition between the one-halo term and the
two-halo term. It has been fully tested against numerical simulations
(\citealt{zehavi_etal:04, tinker_etal:05, chen:07};
Wechsler~et.~al.~2009, in preparation).

The clustering statistic measured by Q08 is the angular correlation
function $\wth$. The three-dimensional correlation function is
connected to the angular clustering by

\begin{equation}
\label{e.wth}
\wth = \int dz\,N^2(z)\, \frac{dr}{dz} \int dx\, \xi\left(\sqrt{x^2 + r^2\theta^2}\right),
\end{equation}

\noindent where $N(z)$ is the normalized redshift distribution of the
galaxy sample, $r$ is the comoving radial distance at redshift $z$ and
$dr/dz = (c/H_0)/\sqrt{\om(1+z)^3+\oml}$. The photometric redshift
distribution is roughly a top-hat function for $2.0<z<2.8$. However,
photometric redshift uncertainties are important in estimating the
true underlying redshift distribution. Unless specified, we use the
estimated true redshift distribution of the sample used by Q08 (kindly
provided by R. Quadri). For this $N(z)$, the redshift distribution
peaks at $z=2.3$ but contains significant wings out to $z\sim 1.5$ and
$z\sim 3.3$ (see Figure 3 in Q08 for comparison). The amplitude of the
$\wth$ is somewhat sensitive to the choice of $N(z)$; for the same
$\xi(r)$, using the photometric $N(z)$ increases the amplitude of
$\wth$ by $\sim 20\%$, independent of scale. Although the uncertainty
in $N(z)$ is a source of error in the theoretical calculations, we
will show that it is subdominant to the sample variance of the
observations themselves. In practice, we calculate $\xi(r)$ at the
peak of $N(z)$, $z=2.3$, and implement that function in equation
(\ref{e.wth}). Q08 recalculate $\xi(r)$ at each $z$ using the
redshift-dependent halo statistics with a fixed HOD. The assumption
that the HOD is fixed in redshift is likely wrong to some degree but
cannot be quantified, thus it is not necessarily more robust to
recalculate $\xi(r)$ as a function of redshift. In practice, we find
that there are negligible differences between the two approaches.

Our motivation for using an analytic model to calculate galaxy
clustering, rather than using one of the simulations, is
two-fold. First, it frees us to use our desired cosmological
model. Second, it ameliorates any numerical issues of spatial
resolution; the Q08 clustering measurements extend down to $\sim 0.02$
\hmpc\ (comoving at $z=2.3$), which is below the limit at which
subhalos can be identified in the simulations. Figure \ref{hod}b
demonstrates the veracity of our analytic model. The points show the
clustering of subhalos in the L120 box while the solid curve shows the
clustering obtained from the analytic model using the HOD in Figure
\ref{hod}a (shifted to the proper mass scale, and using the
cosmology of the L120 box). The good agreement with the
simulation results demonstrate the robustness of our analytic model.

\subsection{Breaking Galaxies into Red and Blue}

DRGs do not comprise a luminosity-threshold sample of galaxies; the
luminosity functions of \cite{marchesini_etal:07} demonstrate that
even at the brightest end, DRGs only account for roughly half of all
galaxies. Our model for the fraction of central galaxies that are DRGs
has a mass-dependence of the form

\begin{equation}
\label{e.fcen}
\fcen(M) = \fmax \exp\left[\frac{-\kappa \mmin}{M-\mmin}\right].
\end{equation}

\noindent Equation (\ref{e.fcen}) implies that all galaxies at
$M=\mmin$ are blue (non-DRGs), and that the fraction of DRGs smoothly
increases to an asymptotic value of $\fmax$ at high masses. If
$\kappa=0$, then $\fcen=\fmax$ at all masses and central DRGs are a
random subsample of all central galaxies. Formally, the parameter
$\kappa$ is the fractional increase in halo mass with respect to
$\mmin$ for $\fcen$ to reach $1/e$ of the asymptotic value $\fmax$.
In lieu of $\kappa$, we will refer to a physically more interesting
quantity, the ratio between the mean halo mass for central DRGs and
the mean halo mass for all central galaxies,

\begin{equation}
\label{e.mshift}
\mp = \frac{ \int_{\mmin}^\infty dM (dn/dM) M \fcen(M) }{\int_{\mmin}^\infty dM (dn/dM) M },
\end{equation}

\noindent where $dn/dM$ is the halo mass function. 

For satellite galaxies we stipulate that a constant fraction $\fsat$
of satellites are DRGs, independent of mass. A more physical model may
have a mass-dependent satellite red fraction, but with the large
errors on the given data, a three-parameter model is a reasonable first
step. 

An important uncertainty in modeling DRG clustering is the uncertainty
in the number density of such objects. Our model for the DRG fraction
has three free parameters ($\kappa$, $\fmax$, and $\fsat$), but
$\ndrg$ reduces the degrees of freedom by one. However, given the
small sample size, the error in $\ndrg$ is non-negligible and must be
taken into account when modeling the data. We will address this in the
following section.

\begin{figure}
\epsscale{1.0} 
\plotone{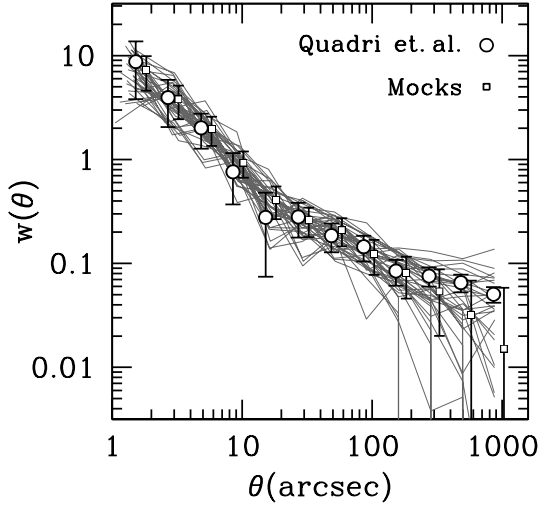}
\caption{ \label{mocks} Dispersion of the clustering from the $15^2$
  mock catalogs. The thin gray curves show $w(\theta)$ for a random
  18\% of the mocks. The open circles are the Q08 data and errors. The
  open squares represent the mean (corrected for the integral
  constraint) and dispersion among the mocks. The squares have been
  shifted slightly for clarity. For all calculations, the covariance
  of the mocks is used as the errors on the Q08 data. }
\end{figure}

\begin{figure}
\epsscale{1.0} 
\plotone{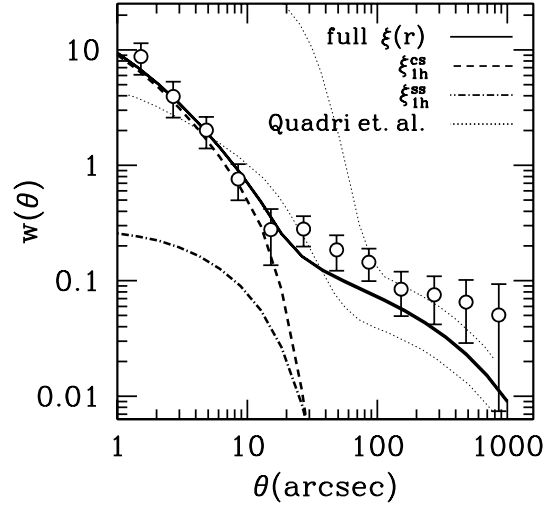}
\caption{ \label{wtheta} Comparison of models to $w(\theta)$
  measurements. The solid curve shows the best-fit HOD model that
  matches the $\wth$ and number density of DRGs. The dashed and
  dashed-dot curves show the breakdown of the one-halo term into
  central-satellite pairs ($\xi_{\rm 1h}^{cs}$) and satellite-satellite
  pairs ($\xi_{\rm 1h}^{ss}$). The thin dotted curves are the two fits
  presented in Q08.}
\end{figure}

\subsection{Error Estimation}

Q08 obtain error bars through bootstrap resampling of their data,
where each bootstrap sample draws from the distribution of all
galaxies. This method is good for estimating errors due to shot noise
but not due to cosmic variance. The measurements of $\wth$ are taken
out to $\sim 1200$ arcsec ($0.33 \deg$), close to half the width of
the field ($0.84 \deg$). Given the volume of the field, roughly
$2.2\times 10^6$ $(h^{-1}$Mpc$)^3$ (about the volume of a ($130$
\hmpc)$^3$ cube), cosmic variance is important for modeling the data
at large-scales.

To estimate the errors on the Q08 data, we construct mock catalogs by
populating the halos in a large-volume simulation (L1000 listed in
Table 1). The simulation itself does not resolve substructure, so we
use the analytic $\navg$ from Figure \ref{hod} to populate the halos
in the simulation. Central galaxies are placed at the center of mass
of the halo and satellite galaxies are placed randomly, assuming a
density profile that follows that of \cite{nfw:96}. Halos are
identified in the simulation using the friends-of-friends technique
(e.g., \citealt{davis_etal:85}). Due to the low mass resolution of the
simulation, we use halos down to 10 particles. This is not ideal, but
the large-scale bias of these halos matches theoretical expectations,
thus the mock-to-mock variations of these halos is accurate
(\citealt{hu_kravtsov:03}). To determine the DRG fraction, we use a
model with parameters $\kappa=0.65$ and $\fsat=0.64$, yielding a value
of $\fmax=0.95$. (This was the best-fit model from fitting the Q08
data using the Q08 error bars and a constant $N(z)$ from $2<z<2.8$.)

We use the simulation output at $z=2.5$.\footnote{Using a single
  redshift output rather than a full lightcone makes negligible
  difference in the clustering. As redshift decreases, the amplitude
  of dark matter clustering increases with the growth factor, but if
  the HOD is fixed (as assumed here and in all HOD fits to $\wth$
  data) then the bias at the mean mass scale of the galaxies decreases
  in such a way as to nearly cancel the change in dark matter
  clustering.} The comoving distance from $2<z<3$ is roughly 1
comoving $\hgpc$, allowing us to fully incorporate the the redshift
depth of the Q08 photometric sample. We use a redshift distribution
that is constant from $z=2$ to $z=2.8$, and zero at higher and lower
redshift. At $z=2.8$, $0.84\deg$ of arc is 64 comoving \hmpc, thus we
are able to create $15^2=225$ mocks.

A random sample of 18\% of the mocks are compared to the Q08 data in
Figure \ref{mocks}. The data has been corrected for the integral
constraint. We also perform the integral constraint correction for the
correlation functions from the mocks by adding a constant. The
constant is computed as the difference between the mean correlation
function of all mocks and the correlation function measured from the
entire box.  The agreement between the mocks and the Q08 data at large
scales is artificially enhanced due to a number of reasons.

The narrow $N(z)$ function increases the amplitude of $\wth$, and with
the smaller error bars at large scales the best-fit model has a lower
number density, yielding a higher bias. The FOF halos with low
particle numbers exhibit somewhat stronger scale-dependence than
that seen in \cite{tinker_etal:09_bias}, which helps the fit at
$\theta\sim 20''$.  However, for the purpose of estimating cosmic
variance the most important criterion is producing mocks the reproduce
the data on most scales.  The large-scale clustering of the mock DRG
samples is consistent with the data given the sample variance at less
than $1-\sigma$. At $\theta\lesssim 50$ arcsec, the bootstrap errors of
Q08 are quite accurate in recovering the error, but the error bar on
the $\theta=10^3$ arcsec datum is underestimated by a factor of
5. Henceforth, we use the dispersion among the mocks as the diagonal
errors on the Q08 data. We note that we use the dispersion as an
absolute error, not a fractional error. Because $\wth$ is a projected
quantity, the data points are highly correlated. Our large number of
mocks allows us to robustly calculate the covariance matrix for the
Q08 sample. For all model fitting we use the covariance matrix. We
also note that Q08 have an additional datum at $\theta=1400$
arcsec. Although the measured value of $\wth$ is significantly lower
than the other data points, we do not use this point because sample
variance renders it essentially useless.

\cite{guo_white:08} use the Millennium semi-analytic model to make
similar mock samples to compare to the Q08 data (a simulation $1/8$
the volume of our L1000 simulation). The dispersion of their 48 mocks
is comparable to Figure \ref{mocks}, but their mean correlation
function is significantly lower than the Q08 data, even at small
scales. Even with this discrepancy, 6 of their 48 mocks match the
large-scale data up to the data at $\theta<10^3$, which is roughly a
1-$\sigma$ result and consistent with our conclusions.

The variance in the number density from the mocks is 11\%, somewhat
better than the 18\% uncertainty on $\ndrg$ from
\citealt{marchesini_etal:07}, owing to the larger sample size. We use
this 11\% uncertainty when modeling the clustering.

\begin{figure*}
\epsscale{1.0} 
\plotone{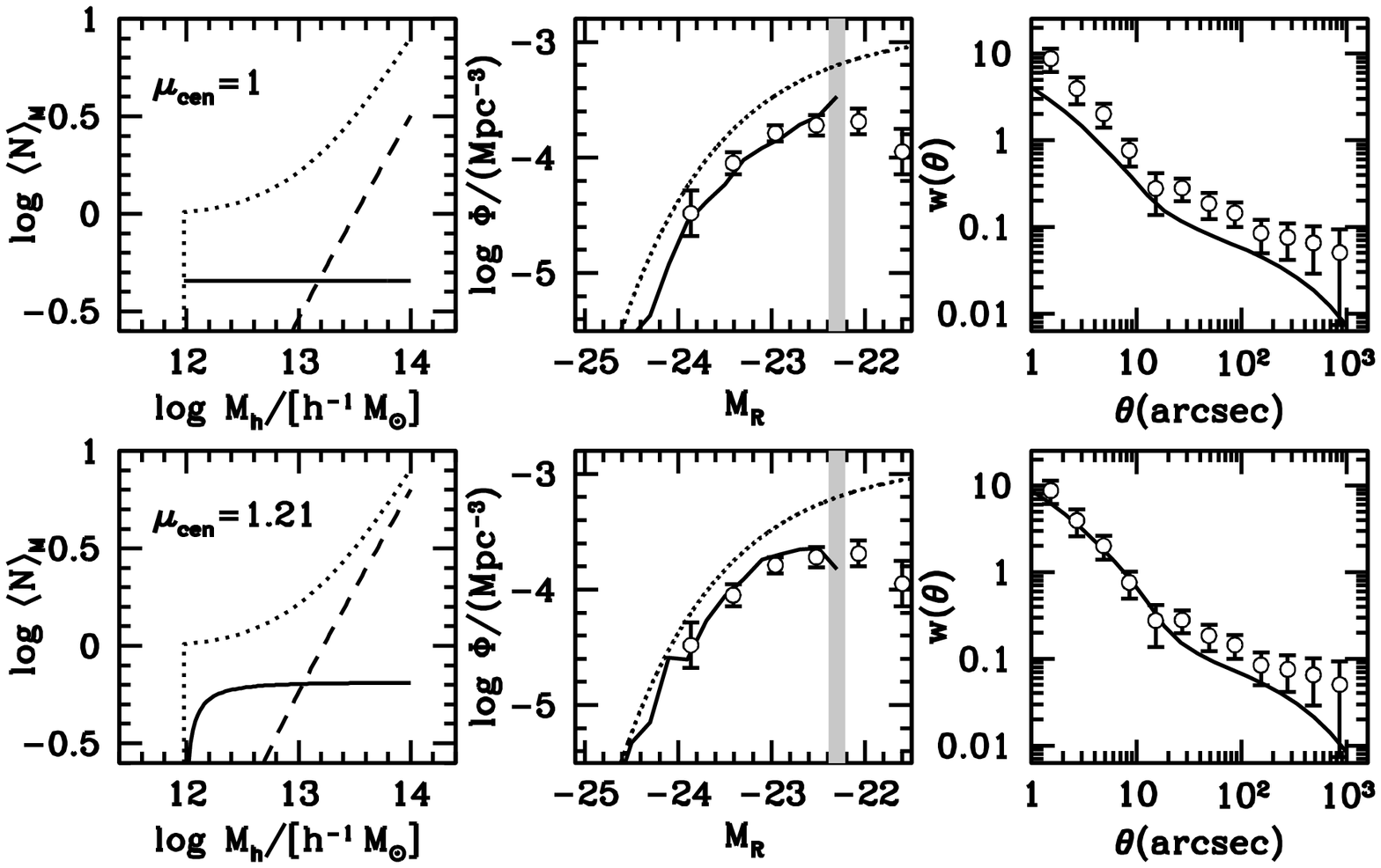}
\vspace{-5.0cm}
\caption{ \label{9panel} Two examples of the combined constraints of
  $w(\theta)$ and $\lfdrg$. The top row shows results from assuming
  that DRGs are a random sample of all galaxies. The top left panel
  shows the HOD; the solid curve shows $\ncen$ for DRGs, and the
  dashed line shows $\nsat$ for DRGs. The dotted curve shows $\navg$
  for all galaxies for comparison. The middle panel shows the
  luminosity function of \cite{marchesini_etal:07} (points with
  errors). The solid line is the model luminosity function. The dotted
  line is the luminosity function for all galaxies. The thick solid
  line is the completeness limit of the Q08 data. The right panel is
  the clustering of DRGs. The points represent the Q08 data, while the
  curve is the model prediction from the HOD in the far left
  panel. {\it Bottom row}: The best-fit model from the combined
  constraints of $\wth$ and $\lfdrg$. The mean mass of central DRGs is
  $21\%$ higher than the overall sample of central galaxies ($\sim
  40\%$ higher than non-DRG central galaxies). This model produces a
  better fit to both the luminosity function and $\wth$. }
\end{figure*}

\section{Results}

\subsection{Fitting the Quadri et.~al. data }

We use a Monte Carlo Markov Chain (MCMC) analysis to determine the
best-fit model parameters as well as their uncertainties. Using $\wth$
and $\ndrg$, the model with the minimum $\chi^2$ has parameters
$\fsat=0.69$, $\fmax=0.99$, and $\kappa=0.48$, yielding a number
density of $6.4\times 10^{-4}$ and $\mp=1.46$. The $\chi^2$ for this
model is 7.0 with $13-3$ degrees of freedom. Figure \ref{wtheta}
compares our best-fit model with the Q08 data. At most scales the
model is within the $1-\sigma$ errors. The highly correlated nature of
the data in the two-halo regime reduces the significance of the offset
between the model and the data. The thin dotted lines show the two
models presented in Q08. The upper curve, which matches the data at
large scales, diverges rapidly from the data at $\theta<100$ arcsec
because nearly all galaxies in this model are satellites in high-mass
halos. The lower curve matches the data adequately on small scales,
but is significantly below the data in the two-halo term, even with
our new errors estimates.

The differences between these two analyses are driven by the treatment
of the second moment at small scales and the choice of halo bias
models at large scales. The calculations of Q08 are based on the model
of \cite{hamana_etal:04} (also used by \citealt{lee_etal:06} to model
the clustering of LBGs, and similar to the model of
\citealt{bullock_etal:02}). In this formulation, the mean occupation
function is a simple power law of the form

\begin{equation}
\label{e.hamana1}
\navg = \left\{ \begin{array}{ll}
  (M/M_1)^{\beta} & {\rm if\ \ } M\ge \mmin \\
  0 & {\rm if\ \ } M< \mmin.\\
\end{array}
\right. 
\end{equation}

\noindent Even though equation (\ref{e.hamana1}) has one fewer free
parameter than our $\navg$ in \S 2.2, it has significantly less
freedom because it does not parameterize centrals and satellites
separately. It is well known that assuming entirely Poisson
fluctuations around $\navg$ produces clustering results that cannot
match observations; the second moment is too high at low $\navg$ (see,
e.g., \citealt{benson_etal:00, seljak:00, roman_etal:01,
  berlind_weinberg:02}). In the HOD model of \S 2, the sub-Poisson
scatter is a natural consequence of parameterizing the centrals and
satellites separately; the second moment of the satellites is always
Poisson, but when $\nsat$ falls below unity the total distribution of
pairs becomes sub-Poisson because there is no scatter (or only a
nearest-integer distribution) of central galaxies (see
\citealt{kravtsov_etal:04, zheng_etal:05}). In addition, the
separation distribution of central-satellite pairs is different than
satellite-satellite pairs. In the former, the pair distribution
follows the halo density profile. In the latter, the pair distribution
is represented by the density profile convolved with itself. In galaxy
distributions with high satellite fractions (say, $\sim 30\%$ for
$L<\lstar$ galaxies), the overall shape of the one-halo term will
differ from samples with a low fraction of satellites (such as LRGs,
DRGs, or any other commensurately bright sample that populates the
high-mass end of the halo mass function; see the discussion in the
Appendix A of \citealt{zheng_etal:08}).

\begin{figure*}
\epsscale{1.15} 
\vspace{-8.5cm}
\plotone{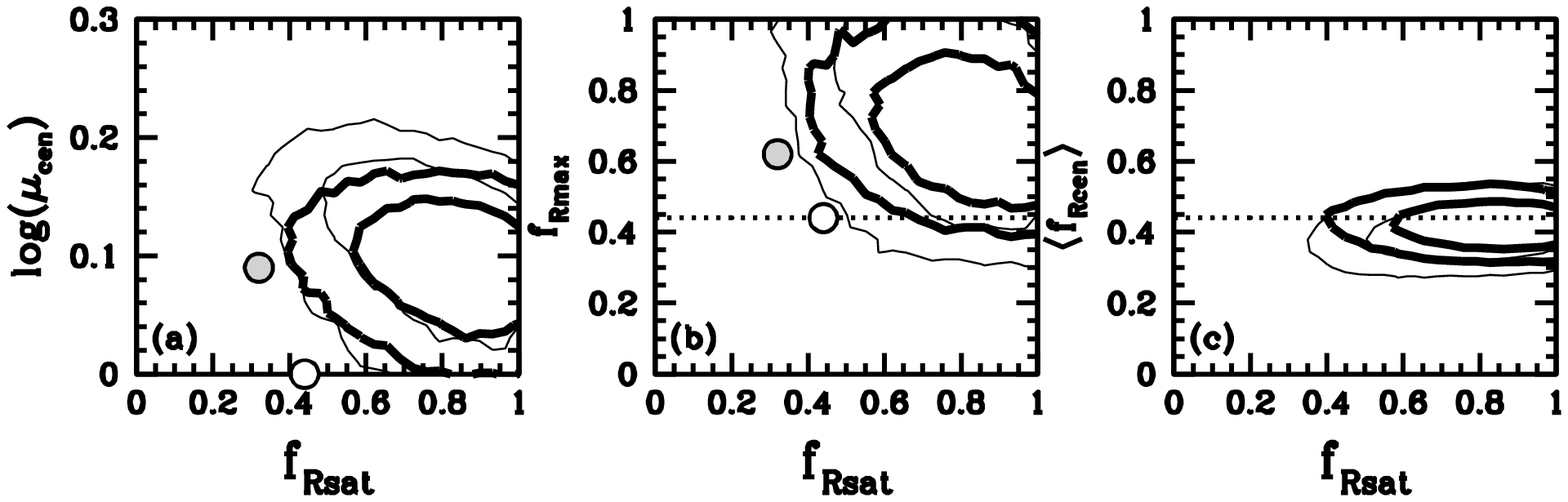}
\caption{ \label{constraints} Panel (a): Constraints on $\fsat$ and
  $\mp$ from the MCMC chains. Thick contours represent the $1-\sigma$
  and $2-\sigma$ constraints using both $\wth$ and $\lfdrg$. Thin
  contours correspond to using $\wth$ alone. The gray dot represents
  the value $\fsat$ obtained if equation (\ref{e.fcen}) were applied
  to the subhalos. The white dot represents a model in which DRGs are
  a random sample of all galaxies. Panel (b): Constraints on $\fsat$
  and $\fmax$. Contours are as in panel (a). The dotted line indicates
  the overall fraction of DRGs, 44\%. Panel (c) shows the constraints
  on $\fsat$ and $\langle f_{\rm cen}\rangle$, the mean central DRG
  fraction integrating over all halos.  }
\end{figure*}

In equation (\ref{e.hamana1}) there is no explicit separation of
central and satellite galaxies. Thus, sub-Poisson fluctuations are
specified in the model {\it ad-hoc} when $\navg<1$. More importantly,
the model specifies a binary switch between radial pair distributions;
at $\navg\ge 1$, the entire one-halo term is calculated assuming a
satellite-satellite profile even though, at that mass scale, almost
{\it all} pairs are still central-satellite. In Figure \ref{wtheta},
the relative contribution of central-satellite pairs, $\xics$, and
satellite-satellite pairs, $\xiss$, to the one-halo term is shown with
the dashed and dot-dashed curves, respectively. At all scales $\xics$
dominates because the overall satellite fraction of this sample is low
($16 \%$); although satellites dominate the pair counts at $M\gtrsim
2M_1$, the number of halos at that mass is nearly negligible because
$M_1$ is on the exponentially-falling tail of the mass function for
this sample. One can see that the shape of the one-halo term from the
\cite{hamana_etal:04} model is the same as $\xiss$. This gives the one
halo term a higher amplitude and larger radial extent than a model in
which the two one-halo terms are properly weighted. Thus, to match at
small scales, the best-fit model forces the large-scale clustering
lower than it otherwise would be.

The differences between Q08 and this work at small scales do not
entirely account for the differences at large scales. Half the
difference can be accounted for by the choice of large-scale bias. Q08
use the bias model of \cite{smt:01} (hereafter SMT), which was
calibrated on small-volume simulations. There is debate in the
literature over the accuracy of the SMT bias function at high masses
(relative to the non-linear mass scale $\mstar$) compared to the
spherical collapse model (e.g., \citealt{cole_kaiser:89,
  mo_white:96}), which predicts stronger bias at these mass scales
(see \citealt{cohn_white:08, reed_etal:08}). Our model is intermediate
between these two models and agrees perfectly with the bias of the
Millennium simulation shown in \cite{gao_etal:05}. Our new bias
prescription increases the large-scale bias of the galaxy sample by
10\% relative to SMT, increasing the large-scale amplitude of $\xi(r)$
by 20\%. The results of Figure \ref{hod}b also show that our bias
prescription is adequately describing the clustering of halos in the
L120 simulation. Using the SMT bias function, our best-fit model has a
$\chi^2=10.0$, while the Q08 model yields $\chi^2=26.4$. Using the SMT
bias function has no effect on the parameter constraints in \S
3.3. For comparison, using diagonal errors only results in $\chi^2$
values of 10.9 for the best-fit model, 19.5 when using SMT bias, and 32.0
for the Q08 model.

\subsection{Including the Luminosity Function}

To a first approximation, rest-frame $R$-band absolute magnitude
should be monotonically correlated with dark matter halo and subhalo
mass. This is borne out by the SHAM results of \cite{conroy_etal:06},
and it is further supported by the comparison between SHAM and the
clustering of $z\sim 2$ star-forming galaxies
(\citealt{conroy_etal:08}). By construction, the luminosity function
of the galaxies in our model exactly follow that measured by
\cite{marchesini_etal:07} for all galaxies (DRG and non-DRG). For each
galactic halo, the luminosity of the galaxy within it is set by
matching the cumulative halo abundance to the cumulative luminosity
function. We can combine the luminosity set in this way with the
mass-dependent DRG occupation function described in \S 2.2 to create a
DRG luminosity function, which has also been measured by
\cite{marchesini_etal:07}. To do this, we use the halos and subhalos
identified in the L160 simulation. Each halo has a luminosity obtained
through the abundance-matching method. Whether or not the galaxy is a
DRG is decided by Monte Carlo based on $\fcen(M)$ and $\fsat$.

Figure \ref{9panel} shows two examples of the combined constraints of
$\wth$, $\ndrg$ and $\lfdrg$\footnote{We include both $\ndrg$ and
  $\lfdrg$ as independent data because $\lfdrg$ does not necessarily
  include $\ndrg$. We find little difference in parameter constraints
  when excluding $\ndrg$.}. In the top row, we show a model in
which DRGs are simply a random sample of all galaxies. Thus $\mp=1$
($\kappa=0$) and $\fmax= \fsat=0.44$. The occupation function for DRGs
is the same as $\navg$ for all galaxies but shifted down by 0.44. The
luminosity function produced by this model is in good agreement with
the observations, but the clustering is clearly inconsistent with the
measurements, both at large and small scales. The bottom row presents
the results from our best-fit model from this combined approach. In
this model, $\kappa=0.18$, $\fsat=0.91$ and $\fmax = 0.66$, yielding
$\mp=1.21$. This shift of the mass scale of central DRGs produces a
noticeable increase in the large-scale bias of $\wth$. The increased
fraction of red satellites increases the amplitude of one-halo term to
agree well with the data. The luminosity function of this model agrees
essentially perfectly with observations, producing the right number of
high-luminosity objects while still matching the faint end down to the
completeness limit of the Q08 sample. With 17-3 degrees of freedom,
this model yields $\chi^2/\nu=7.7/14$.

\subsection{Parameter Constraints}

We can use the observational data to place constraints on $\kappa$,
$\fsat$, and $\fmax$. Figure \ref{constraints} shows the results from
the MCMC chains. The thick contours are the 1- and $2-\sigma$
constraints using $\wth$, $\ndrg$, and $\lfdrg$, while the thin
contours represent results using only $\wth$ and $\ndrg$. Figure
\ref{constraints}a plots the constraints on the $\fsat$-$\mp$
plane. It is clear from Figure \ref{9panel} that a large fraction of
satellites must be red in order to match the clustering; the best-fit
value of $\fsat=0.9$ and the $2-\sigma$ lower limit is $\sim 50\%$ when
marginalizing over other parameters. Ratios of the halo mass scale
between central DRGs and all central galaxies above $\mp=\sim 1.5$ are
excluded at $>2-\sigma$, but a model in which $\mp=1$ (i.e., central
DRGs are a random subsample of all central galaxies) is also excluded
at roughly $2-\sigma$.

Figure \ref{constraints}b shows the parameter constraints in the
$\fsat$-$\fmax$ plane. The dotted line indicates the mean DRG
fraction. This figure highlights the benefit of using the luminosity
function as an additional constraint on the models. When considering
$\wth$ and $\ndrg$ only, the model strives to push the large-scale
bias as high as possible. This results in a best-fit $\fmax$ of unity,
implying all halos above $M\sim 10^{12.5}$ \hmsol\ contain a DRG at
their center. However, this overproduces the number of bright DRGs at
the expense of the low-luminosity end. Using $\wth$ and $\lfdrg$, the
best-fit value of $\fmax$ is 0.66.

Figure \ref{constraints}c shows the constraints on red satellite
fraction and the {\it mean} red central fraction averaged over all
halos. The constraints in all three panels imply that a higher
fraction of satellites than centrals must be red in order to match the
data. (We remind the reader that $\fmax$ is the asymptotic value of
$\fcen(M)$; at fixed $\fsat$, the higher the value of $\fmax$, the
lower the red fraction at lower halo masses in order to keep the
number of central DRGs constant. The overall red central
fraction will usually be near 44\% because the contribution of
satellites is small.) The gray circle in Figures \ref{constraints}a
and \ref{constraints}b shows the $\fsat$ that results if we apply the
mass-dependent $\fcen(M)$ to the subhalos themselves for the best-fit
value of $\kappa$. Because subhalos have a lower mean mass than parent
halos, in this case the fraction of red satellites is less than the fraction
of red centrals if they have the same duty cycle. Thus, using this
model, the fraction of red satellites is 32\%, a value that is clearly
outside the $2-\sigma$ contour.

\begin{figure}
\epsscale{1.0} 
\plotone{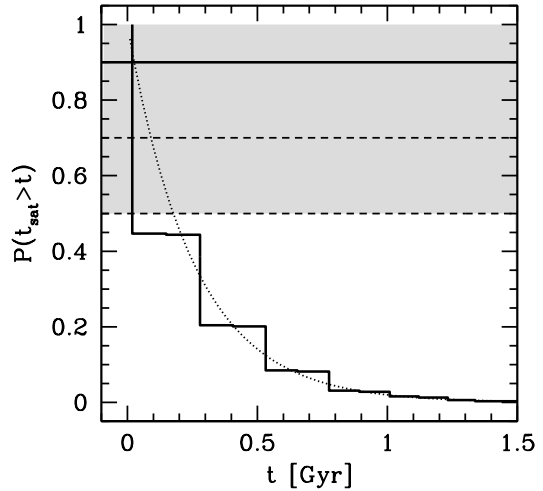}
\caption{ \label{tsat} Cumulative distribution of subhalos with
  accretion times (time since accretion) longer than $t$.  Histogram
  shows the numerical data. The dotted line is a fitting function of
  the form $\exp(-3.9t)$. The shaded region shows the $2-\sigma$
  constraints on the fraction of satellites that are required to be
  DRGs to match the data. The thick solid line is the best-fit value;
  dashed lines represent the 1- and 2-$\sigma$ confidence levels after
  marginalized over other parameters. }
\end{figure}

\section{Discussion}

\subsection{Satellite DRGs}

The results from Figure \ref{constraints} imply that the physical
processes that turn a galaxy red, either by creation of dust or
through star formation history, are different for centrals and
satellites at $z>2$. At low redshift such a scenario is expected;
after accretion, gas is ram-pressure stripped and star formation in
satellites is ``quenched''. However, the situation at high redshift is
quite different; the universe at $z=2.3$ is only $\sim 2.9$ Gyr old in
our cosmology, and most mergers of small halos onto bigger ones have
occurred quite recently. Thus, one would naively expect that red
satellites were red before they were accreted (or would have become
red regardless of accretion); i.e., that the processes that make a
galaxy red are driven primarily by the mass of its galactic halo
rather than their position as an accreted object. Such a scenario is
difficult to reconcile with the Q08 data.

Figure \ref{tsat} shows the cumulative distribution of satellite ages
(i.e., time since they were accreted) for subhalos above the minimum
mass threshold of $\navg$ in the simulations. Because the accretion
times are discretized, the dotted line is a fitting function of the
form $\exp(-3.9t)$ that allows for quick interpolation between the
points. Naively we may infer a quenching time scale based on the
fraction of satellites being DRGs (e.g., 200 Myr from the $2$-$\sigma$
line in Fig. 6) , but this makes the strong assumption that all
satellite DRGs have little to no star formation. Let us assume that
30\% of DRGs have significantly attenuated star formation
(\citealt{labbe_etal:05}), and 32\% of DRG satellites were
classifiably DRGs before accretion (from the gray dot in Figure
\ref{constraints}). Using the best-fit value of $\fsat=0.9$,
$0.9\times(1-0.32)\times0.3=0.18$ is the fraction of all galaxies that
have their star formation attenuated through satellite accretion. From
Figure \ref{tsat}, a star formation quenching timescales of $450$ Myr
is obtained. This is is clear contrast with results from low redshift,
where the exponential timescale for the reduction of star formation in
satellites is closer to $2$-$2.5$ Gyr (\citealt{wang_etal:06,
  kimm_etal:08}). These results can be brought into better agreement
if the exponential timescale scales with the dynamical time of the
galactic halo, reducing the $e$-folding time by $(1+z)^{1.5} = 6$,
depending on the number of $e$-folds required to be considered ``red
and dead''.  Robust conclusions are difficult to establish given the
uncertainty in the fraction of DRGs with no star formation,
uncertainty in how that fraction is distributed between centrals and
satellites, and the uncertainty in $\fsat$.

\begin{figure}
\epsscale{1.0} 
\plotone{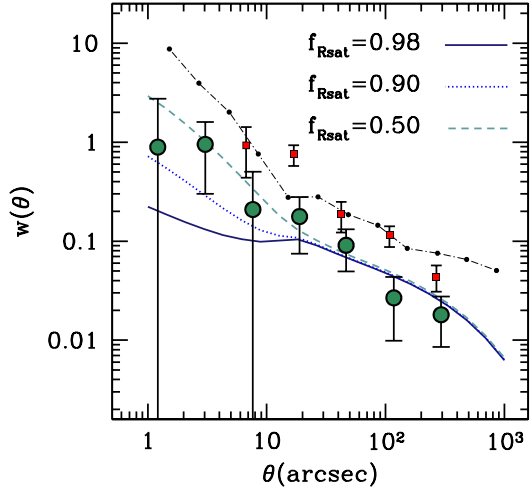}
\caption{ \label{wth_blue} Clustering of blue galaxies (non-DRGs) at
  the same redshift as the Q08 sample. Large circles and their errors
  are taken from \cite{quadri_etal:07}, which show the clustering of
  $J-K<2.3$ galaxies from a smaller sample. The lines are HOD
  predictions for the clustering of non-DRGs from the best-fit value
  of $\mp$ but with varying red satellite fractions. The dotted line is
  the prediction of the overall best-fit model (bottom row of Figure
  \ref{9panel}.) The small squares are the measurement from
  \cite{quadri_etal:07} for DRGs, which are consistent with the Q08
  data (black circles connected by the dash-dot curve). }
\end{figure}

Regardless of current star formation rates, it is also difficult to
reconcile the high overall fraction of satellite DRGs with satellite
accretion timescales. While there is a physical mechanism that halts
star formation for satellite galaxies, accretion onto a larger halo
cannot account for excess $K$-band emission from an advanced
(multi-Gyr) stellar population that is part of the definition of a
DRG. Irrespective of accretion time, the old stars were formed before
the galaxy became a satellite galaxy. Thus the satellite had to know
it would be accreted well beforehand. For $z=2.3$ $10^{13}$ \hmsol\
halos, their average mass was $\sim 1/10$ at $z\sim 5$
(\citealt{wechsler_etal:02}), but their rapid growth and high relative
mass can influence their larger environment and the smaller halos that
will eventually be accreted at later times (\citealt{wang_etal:07,
  hahn_etal:08, dalal_etal:08}). These lower-mass halos within the
lagrangian radius of the large halo will have formed earlier and
possibly have older stellar stellar populations than a halo of the
same mass which formed outside the lagrangian radius of a large
halo. By $z\approx 2.3$, these early-forming halos have become
satellites in the nearby high-mass halos.

It would be perhaps too coincidental that this effect would {\it only}
alter the formation of halos that would become subhalos at the epoch
of the Q08 sample, especially given that most accretion events are
very recent. High mass halos at $z=2.3$ will continue to rapidly grow
and alter the formation trajectories of smaller halos around them,
inducing assembly bias---the correlation between halo occupation and
large-scale environment at fixed halo mass---cited by Q08 to explain
the high clustering amplitude of DRGs. If this assembly bias extends
to halos outside the virial radii of $M\gtrsim 10^{13}$ \hmsol\ halos
(but within their vicinity), the large-scale clustering of DRGs would
be enhanced, supporting the conclusions of Q08. The data do not
require any assembly bias to fit the large scale $\wth$, but it is
implied by the high fraction of satellite galaxies being of DRGs.

Recent developments in stellar population synthesis, specifically the
inclusion of thermally-pulsating AGB stars, may alleviate some of the
tension between the age of the universe at $z=2.3$ and the stellar
population ages required to reproduce the observed colors of DRGs
(\citealt{tonini_etal:08}), although this mechanism is not related to
the high fraction of satellites being DRGs or the fraction of DRGs
without star formation.

\subsection{Central Galaxies}

For central galaxies, the luminosity function alone supports a model
in which DRGs are a random subsample of the halos occupied by bright
galaxies. In contrast, the clustering of DRGs supports a model in
which massive halos ($M\gtrsim 10^{12.5}$) {\it only} house DRGs at
their centers. Considered together, a more intermediate picture
becomes clear in which massive halos host DRGs $\sim 2/3$ of the time,
in good agreement with the mass-selected sample of
\cite{van_dokkum_etal:06}. We have made two strong assumptions when
predicting $\lfdrg$ from our models: 1) that there is no scatter in
the relation between halo mass and luminosity, and 2) that mass and
$R$-band magnitude are monotonically correlated regardless of galaxy
color. For 1), we note that scatter does not change the predicted
$\lfdrg$ when $\mp$ and $\fmax$ are high; these models are excluded
when using $\lfdrg$ (cf, Figure \ref{constraints}b). In such models
{\it all} high-mass halos contain DRGs at their centers, thus scatter
only changes what halo each DRG lives in and not the distribution of
DRG luminosities on the bright end. Significant scatter between halo
mass and luminosity at low luminosities would decrease the observed
amplitude of the correlation function, contrary to the observations
(unless assembly bias is canceling out this effect).  For 2), with the
given data we cannot rule out a model in which red galaxies and blue
galaxies of the same $R$ magnitude occupy halos with substantially
different masses. Our model makes a clear prediction for the
clustering of blue galaxies that is sensitive to this assumption. For
a complete model, the clustering of blue and red galaxies should be
modeled simultaneously.

Figure \ref{wth_blue} shows the clustering of non-DRGs ($J-K<2.3$)
from \cite{quadri_etal:07}, a distinct and much smaller sample of
galaxies than Q08. The three lines are predictions from our best-fit
value of $\kappa=0.18$ for different values of $\fsat$. At large
scales, the measured amplitude of $\wth$ is consistent with our model
predictions. This agreement supports the second assumption above and
argues against significant assembly bias in the DRG population; if the
clustering of DRGs is enhanced by assembly bias then the clustering on
non-DRGs must be suppressed by the same effect. At small scales, there
is a signature of one-halo clustering in the blue galaxies. A model
with nearly no blue satellites appears to be difficult to reconcile
with the observations, but these data cannot distinguish between
models in which 90\% and 50\% of the satellites are red. Because this
is a different galaxy sample from Q08 with a different redshift
distribution, the comparison in Figure \ref{wth_blue} is meant to be
qualitative only. We note, however, that the DRG clustering from
\cite{quadri_etal:07} is consistent with that measured in Q08
(dash-dot curve and squares in Figure \ref{wth_blue}). Measurements of
the clustering of blue galaxies within the Q08 sample will enhanced
our constraints both for halo occupation and for assembly bias. The
relative bias between red and blue galaxies within the same field can
also shed light on the issue of assembly bias. Because they would be
measured from the same volume, large-scale modes would affect the
clustering of both red and blue galaxies in the same way and the
relative bias is therefore independent of cosmic variance
(\citealt{seljak:08, mcdonald_seljak:08}). Assembly bias, on the other
hand, would increase the clustering difference between red and blue
subsamples.

\section{Conclusions}

The clustering of DRGs measured by Q08 can be adequately fit by
standard halo occupation models. With proper estimates of the cosmic
variance at large scales and a more robust implementation of the HOD,
our best-fitting model has $\chi^2/\nu<1$. Thus, from the large-scale
clustering alone there is no compelling evidence for assembly bias in
the halo occupation of DRGs.  The high clustering amplitude of DRGs
has been measured by other authors (\citealt{daddi_etal:03,
  grazian_etal:06}), but the \cite{daddi_etal:03} data is fully
described by HOD modeling as well (\citealt{zheng:04}). Given that the
Q08 sample is by far the largest at present, the inclusion of other
currently available data sets is not likely to change our conclusions,
though cosmic variance is still an important systematic on large
scales.

To fit the prominent break in the $\wth$ between the two-halo and
one-halo terms, a large fraction of satellite galaxies above the
relevant luminosity threshold are required to be DRGs. The best-fit
model has a red satellite fraction, $\fsat$, of nearly unity, with a $2-\sigma$
lower limit of 50\%. This implies that the mechanisms through which
DRGs form are more efficient (or more frequent) for satellite DRGs
than for centrals, even though the accretion times of most subhalos
are very recent at these epochs.  At the best-fit value of
$\fsat=0.9$, the timescale for star-formation quenching is roughly
$450$ Myr if quenching begins at the accretion time. Regardless of the
star formation rates at $z=2.3$, to produce the evolved stellar
populations in satellite DRGs, subhalos are required to know that they
will become subhalos well before they are accreted. This implies these
object have some knowledge of the large-scale environment beyond that
of their host halo mass (i.e.~assembly bias), but further data is
required to quantify the effect.

Down to the completeness limit of the Q08 data, central DRGs have an
average halo mass that is 25\% higher than the overall sample of galaxies
(thus 50\% higher than non-DRGs), but at the highest halo masses DRGs
do not occupy all halos; $\sim 1/3$ of the most massive halos at
$z=2.3$ still contain blue galaxies at their centers.

Clustering data for high-redshift galaxies are becoming precise enough
that realistic models for the occupation of galaxies within halos are
important for robust interpretation.  The approach taken in this work
for modeling halo occupation is more realistic than earlier models
that were developed to modeling high-redshift data
(e.g. \citealt{bullock_etal:02, hamana_etal:04}).  Explicit treatment
of central and satellite galaxies is essential in modeling the
transition regime between one and two halo pairs, which is
particularly important as a constraining feature now that it is well
measured in the data.  The approach taken here, which combines
information from simulations which resolve substructure to constrain
the global sample with a flexible approach for modeling color
dependence, is particularly powerful, and can also be applied to other
color-selected galaxy samples.

\acknowledgments We thank Ryan Quadri, Kyoung-Soo Lee, and Rik
Williams for many helpful discussions. We also thank Martin White for
comments on an earlier version of this draft. Much of the analysis was
performed on the computing facilities at the Kavli Insitute for
Cosmological Physics and the University of Chicago. The N-body
simulations used here were run on the Columbia machine at NASA Ames.
We are grateful to Anatoly Klypin for running these simulations and
making them available to us.  RHW was supported in part by the
U.S. Department of Energy under contract number DE-AC02-76SF00515 and
by a Terman Fellowship at Stanford University. ZZ gratefully
acknowledges support from the Institute for Advanced Study through a
John Bahcall Fellowship.

\appendix
\section{A. Halo Bias}

The large-scale bias of halos used in this paper is based on the
spherical overdensity halo catalogs of \cite{tinker_etal:08_mf}. The
full results for both large-scale bias and scale-dependent bias as a
function of halo overdensity will be given elsewhere
(\citealt{tinker_etal:09_bias}). In the Tinker et.~al.\ results, the
fitting function for large-scale bias is

\begin{equation}
\label{e.bias}
b(\nu) = 1 - A\frac{\nu^{a}}{\nu^{a} + \dc^{a}} + B\nu^{b} + C\nu^{c}.
\end{equation}

\noindent where $\nu=\delta_c/\sigma(M)$, $\delta_c=1.686$ and
$\sigma$ is the linear matter fluctuations within the lagrangian
radius of a halo of mass $M$. For $\Delta=200$ halos, $A=1.04$,
$a=0.132$, $B=0.183$, $b=1.5$, $C=0.262$, and $c=2.4$.


\bibliography{../risa}

\begin{thebibliography}{63}
\expandafter\ifx\csname natexlab\endcsname\relax\def\natexlab#1{#1}\fi

\bibitem[{{Benson} {et~al.}(2000){Benson}, {Cole}, {Frenk}, {Baugh}, \&
  {Lacey}}]{benson_etal:00}
{Benson}, A.~J., {Cole}, S., {Frenk}, C.~S., {Baugh}, C.~M., \& {Lacey}, C.~G.
  2000, \mnras, 311, 793

\bibitem[{{Berlind} \& {Weinberg}(2002)}]{berlind_weinberg:02}
{Berlind}, A.~A. \& {Weinberg}, D.~H. 2002, \apj, 575, 587

\bibitem[{{Bullock} {et~al.}(2002){Bullock}, {Wechsler}, \&
  {Somerville}}]{bullock_etal:02}
{Bullock}, J.~S., {Wechsler}, R.~H., \& {Somerville}, R.~S. 2002, \mnras, 329,
  246

\bibitem[{{Chen}(2007)}]{chen:07}
{Chen}, J. 2007, A\&A, in press, arXiv:0712.0003

\bibitem[{{Cohn} \& {White}(2008)}]{cohn_white:08}
{Cohn}, J.~D. \& {White}, M. 2008, \mnras, 385, 2025

\bibitem[{{Cole} \& {Kaiser}(1989)}]{cole_kaiser:89}
{Cole}, S. \& {Kaiser}, N. 1989, \mnras, 237, 1127

\bibitem[{{Conroy} {et~al.}(2008){Conroy}, {Shapley}, {Tinker}, {Santos}, \&
  {Lemson}}]{conroy_etal:08}
{Conroy}, C., {Shapley}, A.~E., {Tinker}, J.~L., {Santos}, M.~R., \& {Lemson},
  G. 2008, \apj, 679, 1192

\bibitem[{{Conroy} {et~al.}(2006){Conroy}, {Wechsler}, \&
  {Kravtsov}}]{conroy_etal:06}
{Conroy}, C., {Wechsler}, R.~H., \& {Kravtsov}, A.~V. 2006, \apj, 647, 201

\bibitem[{{Cooray} \& {Sheth}(2002)}]{cooray_sheth:02}
{Cooray}, A. \& {Sheth}, R. 2002, \physrep, 372, 1

\bibitem[{{Croton} {et~al.}(2007){Croton}, {Gao}, \& {White}}]{croton_etal:07}
{Croton}, D.~J., {Gao}, L., \& {White}, S.~D.~M. 2007, \mnras, 374, 1303

\bibitem[{{Daddi} {et~al.}(2003){Daddi}, {R{\"o}ttgering}, {Labb{\'e}},
  {Rudnick}, {Franx}, {Moorwood}, {Rix}, {van der Werf}, \& {van
  Dokkum}}]{daddi_etal:03}
{Daddi}, E., {R{\"o}ttgering}, H.~J.~A., {Labb{\'e}}, I., {Rudnick}, G.,
  {Franx}, M., {Moorwood}, A.~F.~M., {Rix}, H.~W., {van der Werf}, P.~P., \&
  {van Dokkum}, P.~G. 2003, \apj, 588, 50

\bibitem[{{Dalal} {et~al.}(2008){Dalal}, {White}, {Bond}, \&
  {Shirokov}}]{dalal_etal:08}
{Dalal}, N., {White}, M., {Bond}, J.~R., \& {Shirokov}, A. 2008, \apj, 687, 12

\bibitem[{{Davis} {et~al.}(1985){Davis}, {Efstathiou}, {Frenk}, \&
  {White}}]{davis_etal:85}
{Davis}, M., {Efstathiou}, G., {Frenk}, C.~S., \& {White}, S. D.~M. 1985, \apj,
  292, 371

\bibitem[{{Dunkley} {et~al.}(2008){Dunkley}, {Komatsu}, {Nolta}, {Spergel},
  {Larson}, {Hinshaw}, {Page}, {Bennett}, {Gold}, {Jarosik}, {Weiland},
  {Halpern}, {Hill}, {Kogut}, {Limon}, {Meyer}, {Tucker}, {Wollack}, \&
  {Wright}}]{dunkley_etal:08}
{Dunkley}, J., {Komatsu}, E., {Nolta}, M.~R., {Spergel}, D.~N., {Larson}, D.,
  {Hinshaw}, G., {Page}, L., {Bennett}, C.~L., {Gold}, B., {Jarosik}, N.,
  {Weiland}, J.~L., {Halpern}, M., {Hill}, R.~S., {Kogut}, A., {Limon}, M.,
  {Meyer}, S.~S., {Tucker}, G.~S., {Wollack}, E., \& {Wright}, E.~L. 2008,
  ApJS, in press

\bibitem[{{F{\"o}rster Schreiber} {et~al.}(2004){F{\"o}rster Schreiber}, {van
  Dokkum}, {Franx}, {Labb{\'e}}, {Rudnick}, {Daddi}, {Illingworth}, {Kriek},
  {Moorwood}, {Rix}, {R{\"o}ttgering}, {Trujillo}, {van der Werf}, {van
  Starkenburg}, \& {Wuyts}}]{forster_etal:04}
{F{\"o}rster Schreiber}, N.~M., {van Dokkum}, P.~G., {Franx}, M., {Labb{\'e}},
  I., {Rudnick}, G., {Daddi}, E., {Illingworth}, G.~D., {Kriek}, M.,
  {Moorwood}, A.~F.~M., {Rix}, H.-W., {R{\"o}ttgering}, H., {Trujillo}, I.,
  {van der Werf}, P., {van Starkenburg}, L., \& {Wuyts}, S. 2004, \apj, 616, 40

\bibitem[{{Franx} {et~al.}(2003){Franx}, {Labb{\'e}}, {Rudnick}, {van Dokkum},
  {Daddi}, {F{\"o}rster Schreiber}, {Moorwood}, {Rix}, {R{\"o}ttgering}, {van
  de Wel}, {van der Werf}, \& {van Starkenburg}}]{franx_etal:03}
{Franx}, M., {Labb{\'e}}, I., {Rudnick}, G., {van Dokkum}, P.~G., {Daddi}, E.,
  {F{\"o}rster Schreiber}, N.~M., {Moorwood}, A., {Rix}, H.-W.,
  {R{\"o}ttgering}, H., {van de Wel}, A., {van der Werf}, P., \& {van
  Starkenburg}, L. 2003, \apjl, 587, L79

\bibitem[{{Gao} {et~al.}(2005){Gao}, {Springel}, \& {White}}]{gao_etal:05}
{Gao}, L., {Springel}, V., \& {White}, S.~D.~M. 2005, MNRAS, 363, L66

\bibitem[{{Gao} \& {White}(2007)}]{gao_white:07}
{Gao}, L. \& {White}, S.~D.~M. 2007, \mnras, 377, L5

\bibitem[{{Grazian} {et~al.}(2006){Grazian}, {Fontana}, {Moscardini},
  {Salimbeni}, {Menci}, {Giallongo}, {de Santis}, {Gallozzi}, {Nonino},
  {Cristiani}, \& {Vanzella}}]{grazian_etal:06}
{Grazian}, A., {Fontana}, A., {Moscardini}, L., {Salimbeni}, S., {Menci}, N.,
  {Giallongo}, E., {de Santis}, C., {Gallozzi}, S., {Nonino}, M., {Cristiani},
  S., \& {Vanzella}, E. 2006, \aap, 453, 507

\bibitem[{{Guo} \& {White}(2008)}]{guo_white:08}
{Guo}, Q. \& {White}, S.~D.~M. 2008, ArXiv e-prints, arXiv:0809.4259 [astro-ph]

\bibitem[{{Hahn} {et~al.}(2008){Hahn}, {Porciani}, {Dekel}, \&
  {Carollo}}]{hahn_etal:08}
{Hahn}, O., {Porciani}, C., {Dekel}, A., \& {Carollo}, C.~M. 2008, \mnras,
  submitted, arXiv:0803.4211 [astro-ph]

\bibitem[{{Hamana} {et~al.}(2004){Hamana}, {Ouchi}, {Shimasaku}, {Kayo}, \&
  {Suto}}]{hamana_etal:04}
{Hamana}, T., {Ouchi}, M., {Shimasaku}, K., {Kayo}, I., \& {Suto}, Y. 2004,
  \mnras, 347, 813

\bibitem[{{Hu} \& {Kravtsov}(2003)}]{hu_kravtsov:03}
{Hu}, W. \& {Kravtsov}, A.~V. 2003, \apj, 584, 702

\bibitem[{{Ichikawa} {et~al.}(2007){Ichikawa}, {Suzuki}, {Tokoku}, {Uchimoto},
  {Konishi}, {Yoshikawa}, {Kajisawa}, {Ouchi}, {Hamana}, {Akiyama},
  {Nishimura}, {Omata}, {Tanaka}, \& {Yamada}}]{ichikawa_etal:07}
{Ichikawa}, T., {Suzuki}, R., {Tokoku}, C., {Uchimoto}, Y.~K., {Konishi}, M.,
  {Yoshikawa}, T., {Kajisawa}, M., {Ouchi}, M., {Hamana}, T., {Akiyama}, M.,
  {Nishimura}, T., {Omata}, K., {Tanaka}, I., \& {Yamada}, T. 2007, \pasj, 59,
  1081

\bibitem[{{Kimm} {et~al.}(2008){Kimm}, {Somerville}, {Yi}, {van den Bosch},
  {Salim}, {Fontanot}, {Monaco}, {Mo}, {Pasquali}, {Rich}, \&
  {Yang}}]{kimm_etal:08}
{Kimm}, T., {Somerville}, R.~S., {Yi}, S.~K., {van den Bosch}, F.~C., {Salim},
  S., {Fontanot}, F., {Monaco}, P., {Mo}, H., {Pasquali}, A., {Rich}, R.~M., \&
  {Yang}, X. 2008, ArXiv e-prints, arXiv:0810.2794 [astro-ph]

\bibitem[{{Kochanek} {et~al.}(2003){Kochanek}, {White}, {Huchra}, {Macri},
  {Jarrett}, {Schneider}, \& {Mader}}]{kochanek_etal:03}
{Kochanek}, C.~S., {White}, M., {Huchra}, J., {Macri}, L., {Jarrett}, T.~H.,
  {Schneider}, S.~E., \& {Mader}, J. 2003, \apj, 585, 161

\bibitem[{{Kravtsov} {et~al.}(2004){Kravtsov}, {Berlind}, {Wechsler}, {Klypin},
  {Gottl{\" o}ber}, {Allgood}, \& {Primack}}]{kravtsov_etal:04}
{Kravtsov}, A.~V., {Berlind}, A.~A., {Wechsler}, R.~H., {Klypin}, A.~A.,
  {Gottl{\" o}ber}, S., {Allgood}, B., \& {Primack}, J.~R. 2004, \apj, 609, 35

\bibitem[{{Kravtsov} {et~al.}(1997){Kravtsov}, {Klypin}, \&
  {Khokhlov}}]{kravtsov_etal:97}
{Kravtsov}, A.~V., {Klypin}, A.~A., \& {Khokhlov}, A.~M. 1997, ApJS, 111, 73

\bibitem[{{Kriek} {et~al.}(2006){Kriek}, {van Dokkum}, {Franx}, {Quadri},
  {Gawiser}, {Herrera}, {Illingworth}, {Labb{\'e}}, {Lira}, {Marchesini},
  {Rix}, {Rudnick}, {Taylor}, {Toft}, {Urry}, \& {Wuyts}}]{kriek_etal:06}
{Kriek}, M., {van Dokkum}, P.~G., {Franx}, M., {Quadri}, R., {Gawiser}, E.,
  {Herrera}, D., {Illingworth}, G.~D., {Labb{\'e}}, I., {Lira}, P.,
  {Marchesini}, D., {Rix}, H.-W., {Rudnick}, G., {Taylor}, E.~N., {Toft}, S.,
  {Urry}, C.~M., \& {Wuyts}, S. 2006, \apjl, 649, L71

\bibitem[{{Labb{\'e}} {et~al.}(2005){Labb{\'e}}, {Huang}, {Franx}, {Rudnick},
  {Barmby}, {Daddi}, {van Dokkum}, {Fazio}, {Schreiber}, {Moorwood}, {Rix},
  {R{\"o}ttgering}, {Trujillo}, \& {van der Werf}}]{labbe_etal:05}
{Labb{\'e}}, I., {Huang}, J., {Franx}, M., {Rudnick}, G., {Barmby}, P.,
  {Daddi}, E., {van Dokkum}, P.~G., {Fazio}, G.~G., {Schreiber}, N.~M.~F.,
  {Moorwood}, A.~F.~M., {Rix}, H.-W., {R{\"o}ttgering}, H., {Trujillo}, I., \&
  {van der Werf}, P. 2005, \apjl, 624, L81

\bibitem[{{Lee} {et~al.}(2006){Lee}, {Giavalisco}, {Gnedin}, {Somerville},
  {Ferguson}, {Dickinson}, \& {Ouchi}}]{lee_etal:06}
{Lee}, K.-S., {Giavalisco}, M., {Gnedin}, O.~Y., {Somerville}, R.~S.,
  {Ferguson}, H.~C., {Dickinson}, M., \& {Ouchi}, M. 2006, \apj, 642, 63

\bibitem[{{Lin} {et~al.}(2004){Lin}, {Mohr}, \& {Stanford}}]{lin_etal:04}
{Lin}, Y.-T., {Mohr}, J.~J., \& {Stanford}, S.~A. 2004, \apj, 610, 745

\bibitem[{{Marchesini} {et~al.}(2007){Marchesini}, {van Dokkum}, {Quadri},
  {Rudnick}, {Franx}, {Lira}, {Wuyts}, {Gawiser}, {Christlein}, \&
  {Toft}}]{marchesini_etal:07}
{Marchesini}, D., {van Dokkum}, P., {Quadri}, R., {Rudnick}, G., {Franx}, M.,
  {Lira}, P., {Wuyts}, S., {Gawiser}, E., {Christlein}, D., \& {Toft}, S. 2007,
  \apj, 656, 42

\bibitem[{{Mar{\'{\i}}n} {et~al.}(2008){Mar{\'{\i}}n}, {Wechsler}, {Frieman},
  \& {Nichol}}]{marin_etal:08}
{Mar{\'{\i}}n}, F.~A., {Wechsler}, R.~H., {Frieman}, J.~A., \& {Nichol}, R.~C.
  2008, \apj, 672, 849

\bibitem[{{McDonald} \& {Seljak}(2008)}]{mcdonald_seljak:08}
{McDonald}, P. \& {Seljak}, U. 2008, ArXiv e-prints, arXiv:0810.0323 [astro-ph]

\bibitem[{{Mo} \& {White}(1996)}]{mo_white:96}
{Mo}, H.~J. \& {White}, S.~D.~M. 1996, \mnras, 282, 347

\bibitem[{{Navarro} {et~al.}(1996){Navarro}, {Frenk}, \& {White}}]{nfw:96}
{Navarro}, J.~F., {Frenk}, C.~S., \& {White}, S. D.~M. 1996, \apj, 462, 563

\bibitem[{{Quadri} {et~al.}(2007){Quadri}, {van Dokkum}, {Gawiser}, {Franx},
  {Marchesini}, {Lira}, {Rudnick}, {Herrera}, {Maza}, {Kriek}, {Labb{\'e}}, \&
  {Francke}}]{quadri_etal:07}
{Quadri}, R., {van Dokkum}, P., {Gawiser}, E., {Franx}, M., {Marchesini}, D.,
  {Lira}, P., {Rudnick}, G., {Herrera}, D., {Maza}, J., {Kriek}, M.,
  {Labb{\'e}}, I., \& {Francke}, H. 2007, \apj, 654, 138

\bibitem[{{Quadri} {et~al.}(2008){Quadri}, {Williams}, {Lee}, {Franx}, {van
  Dokkum}, \& {Brammer}}]{quadri_etal:08}
{Quadri}, R.~F., {Williams}, R.~J., {Lee}, K.-S., {Franx}, M., {van Dokkum},
  P., \& {Brammer}, G.~B. 2008, \apjl, 685, L1

\bibitem[{{Reed} {et~al.}(2008){Reed}, {Bower}, {Frenk}, {Jenkins}, \&
  {Theuns}}]{reed_etal:08}
{Reed}, D.~S., {Bower}, R., {Frenk}, C.~S., {Jenkins}, A., \& {Theuns}, T.
  2008, ArXiv e-prints, arXiv:0804.0004 [astro-ph]

\bibitem[{{Scoccimarro} {et~al.}(2001){Scoccimarro}, {Sheth}, {Hui}, \&
  {Jain}}]{roman_etal:01}
{Scoccimarro}, R., {Sheth}, R.~K., {Hui}, L., \& {Jain}, B. 2001, \apj, 546, 20

\bibitem[{{Seljak}(2000)}]{seljak:00}
{Seljak}, U. 2000, \mnras, 318, 203

\bibitem[{{Seljak}(2008)}]{seljak:08}
---. 2008, ArXiv e-prints, arXiv:0807.1770 [astro-ph]

\bibitem[{{Sheth} {et~al.}(2001){Sheth}, {Mo}, \& {Tormen}}]{smt:01}
{Sheth}, R.~K., {Mo}, H.~J., \& {Tormen}, G. 2001, \mnras, 323, 1

\bibitem[{{Spergel} {et~al.}(2007){Spergel}, {Bean}, {Dor{\'e}}, {Nolta},
  {Bennett}, {Dunkley}, {Hinshaw}, {Jarosik}, {Komatsu}, {Page}, {Peiris},
  {Verde}, {Halpern}, {Hill}, {Kogut}, {Limon}, {Meyer}, {Odegard}, {Tucker},
  {Weiland}, {Wollack}, \& {Wright}}]{spergel_etal:07}
{Spergel}, D.~N., {Bean}, R., {Dor{\'e}}, O., {Nolta}, M.~R., {Bennett}, C.~L.,
  {Dunkley}, J., {Hinshaw}, G., {Jarosik}, N., {Komatsu}, E., {Page}, L.,
  {Peiris}, H.~V., {Verde}, L., {Halpern}, M., {Hill}, R.~S., {Kogut}, A.,
  {Limon}, M., {Meyer}, S.~S., {Odegard}, N., {Tucker}, G.~S., {Weiland},
  J.~L., {Wollack}, E., \& {Wright}, E.~L. 2007, \apjs, 170, 377

\bibitem[{{Spergel} {et~al.}(2003){Spergel}, {Verde}, {Peiris}, {Komatsu},
  {Nolta}, {Bennett}, {Halpern}, {Hinshaw}, {Jarosik}, {Kogut}, {Limon},
  {Meyer}, {Page}, {Tucker}, {Weiland}, {Wollack}, \&
  {Wright}}]{spergel_etal:03}
{Spergel}, D.~N., {Verde}, L., {Peiris}, H.~V., {Komatsu}, E., {Nolta}, M.~R.,
  {Bennett}, C.~L., {Halpern}, M., {Hinshaw}, G., {Jarosik}, N., {Kogut}, A.,
  {Limon}, M., {Meyer}, S.~S., {Page}, L., {Tucker}, G.~S., {Weiland}, J.~L.,
  {Wollack}, E., \& {Wright}, E.~L. 2003, \apjs, 148, 175

\bibitem[{{Tinker} {et~al.}(2008{\natexlab{a}}){Tinker}, {Kravtsov}, {Klypin},
  {Abazajian}, {Warren}, {Yepes}, {Gottl{\"o}ber}, \&
  {Holz}}]{tinker_etal:08_mf}
{Tinker}, J., {Kravtsov}, A.~V., {Klypin}, A., {Abazajian}, K., {Warren}, M.,
  {Yepes}, G., {Gottl{\"o}ber}, S., \& {Holz}, D.~E. 2008{\natexlab{a}}, \apj,
  688, 709

\bibitem[{{Tinker} {et~al.}(2008{\natexlab{b}}){Tinker}, {Conroy}, {Norberg},
  {Patiri}, {Weinberg}, \& {Warren}}]{tinker_etal:08_voids}
{Tinker}, J.~L., {Conroy}, C., {Norberg}, P., {Patiri}, S.~G., {Weinberg},
  D.~H., \& {Warren}, M.~S. 2008{\natexlab{b}}, \apj, 686, 53

\bibitem[{{Tinker} {et~al.}(2005){Tinker}, {Weinberg}, {Zheng}, \&
  {Zehavi}}]{tinker_etal:05}
{Tinker}, J.~L., {Weinberg}, D.~H., {Zheng}, Z., \& {Zehavi}, I. 2005, \apj,
  631, 41

\bibitem[{{Tinker} {et~al.}(2009)}]{tinker_etal:09_bias}
{Tinker}, J.~T. {et~al.} 2009, in preparation

\bibitem[{{Tonini} {et~al.}(2008){Tonini}, {Maraston}, {Devriendt}, {Thomas},
  \& {Silk}}]{tonini_etal:08}
{Tonini}, C., {Maraston}, C., {Devriendt}, J., {Thomas}, D., \& {Silk}, J.
  2008, \mnras, submitted, arXiv:0812.1225 [astro-ph]

\bibitem[{{van Dokkum} {et~al.}(2003){van Dokkum}, {F{\"o}rster Schreiber},
  {Franx}, {Daddi}, {Illingworth}, {Labb{\'e}}, {Moorwood}, {Rix},
  {R{\"o}ttgering}, {Rudnick}, {van der Wel}, {van der Werf}, \& {van
  Starkenburg}}]{van_dokkum_etal:03}
{van Dokkum}, P.~G., {F{\"o}rster Schreiber}, N.~M., {Franx}, M., {Daddi}, E.,
  {Illingworth}, G.~D., {Labb{\'e}}, I., {Moorwood}, A., {Rix}, H.-W.,
  {R{\"o}ttgering}, H., {Rudnick}, G., {van der Wel}, A., {van der Werf}, P.,
  \& {van Starkenburg}, L. 2003, \apjl, 587, L83

\bibitem[{{van Dokkum} {et~al.}(2006){van Dokkum}, {Quadri}, {Marchesini},
  {Rudnick}, {Franx}, {Gawiser}, {Herrera}, {Wuyts}, {Lira}, {Labb{\'e}},
  {Maza}, {Illingworth}, {F{\"o}rster Schreiber}, {Kriek}, {Rix}, {Taylor},
  {Toft}, {Webb}, \& {Yi}}]{van_dokkum_etal:06}
{van Dokkum}, P.~G., {Quadri}, R., {Marchesini}, D., {Rudnick}, G., {Franx},
  M., {Gawiser}, E., {Herrera}, D., {Wuyts}, S., {Lira}, P., {Labb{\'e}}, I.,
  {Maza}, J., {Illingworth}, G.~D., {F{\"o}rster Schreiber}, N.~M., {Kriek},
  M., {Rix}, H.-W., {Taylor}, E.~N., {Toft}, S., {Webb}, T., \& {Yi}, S.~K.
  2006, \apjl, 638, L59

\bibitem[{{Wang} {et~al.}(2007){Wang}, {Mo}, \& {Jing}}]{wang_etal:07}
{Wang}, H.~Y., {Mo}, H.~J., \& {Jing}, Y.~P. 2007, \mnras, 375, 633

\bibitem[{{Wang} {et~al.}(2006){Wang}, {Li}, {Kauffmann}, \& {De
  Lucia}}]{wang_etal:06}
{Wang}, L., {Li}, C., {Kauffmann}, G., \& {De Lucia}, G. 2006, \mnras, 371, 537

\bibitem[{{Wechsler} {et~al.}(2002){Wechsler}, {Bullock}, {Primack},
  {Kravtsov}, \& {Dekel}}]{wechsler_etal:02}
{Wechsler}, R.~H., {Bullock}, J.~S., {Primack}, J.~R., {Kravtsov}, A.~V., \&
  {Dekel}, A. 2002, \apj, 568, 52

\bibitem[{{Wechsler} {et~al.}(2006){Wechsler}, {Zentner}, {Bullock},
  {Kravtsov}, \& {Allgood}}]{wechsler_etal:06}
{Wechsler}, R.~H., {Zentner}, A.~R., {Bullock}, J.~S., {Kravtsov}, A.~V., \&
  {Allgood}, B. 2006, \apj, 652, 71

\bibitem[{{Wetzel} {et~al.}(2008){Wetzel}, {Cohn}, \& {White}}]{wetzel_etal:08}
{Wetzel}, A.~R., {Cohn}, J.~D., \& {White}, M. 2008, ArXiv e-prints

\bibitem[{{Zehavi} {et~al.}(2004){Zehavi}, {Weinberg}, {Zheng}, {Berlind},
  {Frieman}, {Scoccimarro}, {Sheth}, {Blanton}, {Tegmark}, {Mo},
  {et~al.}}]{zehavi_etal:04}
{Zehavi}, I., {Weinberg}, D.~H., {Zheng}, Z., {Berlind}, A.~A., {Frieman},
  J.~A., {Scoccimarro}, R., {Sheth}, R.~K., {Blanton}, M.~R., {Tegmark}, M.,
  {Mo}, H.~J., {et~al.} 2004, \apj, 608, 16

\bibitem[{{Zhao} {et~al.}(2008){Zhao}, {Jing}, {Mo}, \&
  {Boerner}}]{zhao_etal:08}
{Zhao}, D.~H., {Jing}, Y.~P., {Mo}, H.~J., \& {Boerner}, G. 2008, ArXiv
  e-prints

\bibitem[{{Zheng}(2004)}]{zheng:04}
{Zheng}, Z. 2004, \apj, 610, 61

\bibitem[{{Zheng} {et~al.}(2005){Zheng}, {Berlind}, {Weinberg}, {Benson},
  {Baugh}, {Cole}, {Dav{\'e}}, {Frenk}, {Katz}, \& {Lacey}}]{zheng_etal:05}
{Zheng}, Z., {Berlind}, A.~A., {Weinberg}, D.~H., {Benson}, A.~J., {Baugh},
  C.~M., {Cole}, S., {Dav{\'e}}, R., {Frenk}, C.~S., {Katz}, N., \& {Lacey},
  C.~G. 2005, \apj, 633, 791

\bibitem[{{Zheng} {et~al.}(2008){Zheng}, {Zehavi}, {Eisenstein}, {Weinberg}, \&
  {Jing}}]{zheng_etal:08}
{Zheng}, Z., {Zehavi}, I., {Eisenstein}, D.~J., {Weinberg}, D.~H., \& {Jing},
  Y. 2008, \apj, submitted, arXiv:0809.1868 [astro-ph]

\end{thebibliography}

\end{document}